\documentclass[reprint,
superscriptaddress,
bibnotes,
 amsmath,amssymb,
 aip,
apl,
showkeys,
floatfix,
]{revtex4-1}

\usepackage{placeins}
\usepackage{graphicx}
\usepackage{dcolumn}
\usepackage[T1]{fontenc}
\usepackage[utf8]{inputenc}
\usepackage{graphicx}
\usepackage{float}
\usepackage{natbib}
\usepackage{hyphenat}
\usepackage{tabularx}
\usepackage{amsmath}
\usepackage{adjustbox}
\usepackage{longtable}
\usepackage{epstopdf}
\usepackage{multirow}
\DeclareUnicodeCharacter{0301}{\'{e}}
\usepackage{bm}

\begin{document}

\preprint{APS/123-QED}

\title{Magnon mediated spin pumping by coupled ferrimagnetic garnets heterostructure}

\author{Anupama Swain*}
%\email{swainanupama@niser.ac.in}
\affiliation{Laboratory for Nanomagnetism and Magnetic Materials (LNMM), School of Physical Sciences, National Institute of Science Education and Research (NISER), An OCC of Homi Bhabha National Institute (HBNI), Jatni-752050, Odisha, India}

\author{Kshitij Singh Rathore*}
%\email{Kshitij.rathore@niser.ac.in}
\affiliation{Laboratory for Nanomagnetism and Magnetic Materials (LNMM), School of Physical Sciences, National Institute of Science Education and Research (NISER), An OCC of Homi Bhabha National Institute (HBNI), Jatni-752050, Odisha, India}

\author{Pushpendra Gupta}
%\email{pushpendra.gupta@niser.ac.in}
\affiliation{Laboratory for Nanomagnetism and Magnetic Materials (LNMM), School of Physical Sciences, National Institute of Science Education and Research (NISER), An OCC of Homi Bhabha National Institute (HBNI), Jatni-752050, Odisha, India}

\author{Abhisek Mishra}
%\email{abhisek.mishra@niser.ac.in }
\affiliation{Laboratory for Nanomagnetism and Magnetic Materials (LNMM), School of Physical Sciences, National Institute of Science Education and Research (NISER), An OCC of Homi Bhabha National Institute (HBNI), Jatni-752050, Odisha, India}

\author{Gary Lee}
%\email{yhlee@u.nus.edu}
\affiliation{Department of Physics, 2 Science Drive 3, National University of Singapore, 117551, Republic of Singapore}

\author{Jinho Lim}
%\email{jly383@illinois.edu}
\affiliation{Department of Materials Science and Engineering and Materials Research Laboratory, University of Illinois Urbana-Champaign, Urbana, Illinois 61801, USA}

\author{Axel Hoffmann}
%\email{axelh@illinois.edu}
\affiliation{Department of Materials Science and Engineering and Materials Research Laboratory, University of Illinois Urbana-Champaign, Urbana, Illinois 61801, USA}

\author{Ramanathan Mahendiran}
%\email{phyrm@nus.edu.sg}
\affiliation{Department of Physics, 2 Science Drive 3, National University of Singapore, 117551, Republic of Singapore}

\author{Subhankar Bedanta}
\email{sbedanta@niser.ac.in}
\affiliation{Laboratory for Nanomagnetism and Magnetic Materials (LNMM), School of Physical Sciences, National Institute of Science Education and Research (NISER), An OCC of Homi Bhabha National Institute (HBNI), Jatni-752050, Odisha, India}
\affiliation{Center for Interdisciplinary Sciences (CIS), NISER, An OCC of Homi Bhabha National Institute (HBNI), Jatni-752050, Odisha, India\\
*equal contribution}

\begin{abstract}
\section*{ABSTRACT}

Spin pumping has significant implications for spintronics, providing a mechanism to manipulate and transport spins for information processing. Understanding and harnessing spin currents through spin pumping is critical for the development of efficient spintronic devices. The use of a magnetic insulator with low damping, enhances the signal-to-noise ratio in crucial experiments such as spin-torque ferromagnetic resonance (FMR) and spin pumping. A magnetic insulator coupled with a heavy metal or quantum material offers a more straight forward model system, especially when investigating spin-charge interconversion processes to greater accuracy. This simplicity arises from the absence of unwanted effects caused by conduction electrons unlike in ferromagnetic metals. Here, we investigate the spin pumping in coupled ferrimagnetic (FiM) Y$_{3}$Fe$_{5}$O$_{12}$ (YIG)/Tm$_{3}$Fe$_{5}$O$_{12}$ (TmIG) bilayers combined with heavy-metal (Pt) using the inverse spin Hall effect (ISHE). It is observed that magnon transmission occurs at both of the FiMs FMR positions. The enhancement of spin pumping voltage ($V_{sp}$) in the FiM garnet heterostructures is attributed to the strong interfacial exchange coupling between FiMs. The modulation of $V_{sp}$ is achieved by tuning the bilayer structure. Further, the spin mixing conductance for these coupled systems is found to be $\approx10^{18}$ m$^{-2}$. Our findings describe a novel coupled FiM system for the investigation of magnon coupling providing new prospects for magnonic devices.

\end{abstract}


\maketitle
The need for ultra-low power consumption devices has given rise to the field of spintronics. This field uses the spin degree of freedom of electrons\cite{bader2010spintronics,vzutic2004spintronics}. Spintronic-based applications rely on the generation of spin currents and their conversion into charge currents in magnetic heterostructures \cite{chumak2015magnon,singh2021high,gupta2021simultaneous,thiruvengadam2022anisotropy}. However, in this process, the scattering of conduction electrons in the magnetic layer results in Joule heating. In this context, moving towards the magnon-based information processing in insulators will solve this issue as there is no physical movement of the electrons\cite{barman20212021}. Magnons are the quanta of spin waves, defined as the collective excitations of magnetic moments in magnetically ordered 
materials. Magnons can propagate over distances ranging from micrometers in low damping metallic thin films to around cm in high quality magnetic insulators \cite{kruglyak2010magnonics,pirro2021advances}. The magnonic spin current can be employed to carry, transport, and process information\cite{costache2006electrical,saitoh2006conversion}, as well as generate a spin torque acting on the local magnetic moment that can be exploited to drive magnetization dynamics\cite{sandweg2011spin,kurebayashi2011spin} and magnetic domain walls\cite{tashiro2012thickness,jungfleisch2011temporal,mee1967rh}. A promising technique for detecting magnonic spin currents is the spin pumping induced inverse spin Hall effect (ISHE) \cite{costache2006electrical,roy2022spin}. Spin pumping refers to the transfer of spin angular momentum via magnetization precession from the ferromagnetic material to the adjacent spin sink layer\cite{mishra2024spin}. These pure spin currents are transformed into conventional charge currents by the ISHE, which allows for a convenient electrical detection of spin-wave based spin currents \cite{saitoh2006conversion}. After the discovery of the spin-pumping effect and the ways for enhancement of spin current in ferrimagnetic insulator (yttrium iron garnet, Y${_3}$Fe$_{5}$O$_{12}$, YIG)/non-magnetic metal (platinum, Pt), there was rapidly emerging interest in the investigation of these structures \cite{chumak2012direct,castera1984state}. In these magnetic insulators movement of individual electron gets restricted, which helps to avoid Joule heating dissipation and therefore benefits modern upcoming spintronic devices\cite{wu2013recent}.

YIG is often considered as the best medium for spin wave  propagation because of its very small Gilbert damping coefficient ($2\times10^{-5}$ for bulk YIG) \cite{hauser2016yttrium}. Being an electrical insulator, electron-mediated angular momentum transfer can only occur at the interface between YIG and a metallic layer. In that context, metals with large spin orbit coupling (SOC) like Pt where a pure spin current can be generated through spin Hall effect (SHE) have been used to excite or amplify propagating spin waves through loss compensation in YIG\cite{hirsch1999spin,wang2011control,padron2011amplification}. Recently, other insulator garnets like Tm${_3}$Fe$_{5}$O$_{12}$ (TmIG), Gd${_3}$Fe$_{5}$O$_{12}$ (GdIG) etc. have been investigated with a focus on low Gilbert damping, interlayer coupling and spintronics applications.
Here we investigate the manipulation, generation and detection of magnon-based spin currents in ferrimagnetic insulators capped with heavy metal Pt. These insulators with Pt have been explored a lot for magnon transport physics in effects like ISHE\cite{haertinger2015spin}, Spin Hall magnetoresistance (SMR)\cite{chen2016theory} and spin-Seebeck effect (SSE) \cite{uchida2010observation}. There are few results published on spin waves in coupled ferrimagnetic layers \cite{baker2016spin,timopheev2014dynamic}, however, the effect of coupling on spin pumping in such FiM system is scarce in literature. In order to study the interface and the growth effects on spin pumping, this work presents a systematic study by considering YIG/Pt, TmIG/Pt and bilayers of YIG/TmIG with Pt. This study reveals the increase in the spin pumping voltage in the bilayers which can be attributed to the interfacial exchange coupling between YIG and TmIG. 

High-quality garnet films were prepared on (111) oriented $\mathrm{GGG}\left(\mathrm{GdGa}_5 \mathrm{O}_{12}\right)$ single crystal substrate by pulsed laser deposition (PLD) technique using an excimer laser $(\lambda=248 \mathrm{~nm}$). The garnet targets were commercially purchased from $\mathrm{M} / \mathrm{s}$. Testbourne, UK. The base pressure of the chamber was $8 \times 10^{-7}$ mbar. The substrate temperature was maintained at 560 $^{\circ} \mathrm{C}$ in a $9 \times 10^{-4}$ mbar of oxygen partial pressure during YIG layer deposition. The laser fluence and repetition rate were $1.8 \mathrm{~J} / \mathrm{cm}^2$ and $8 \mathrm{~Hz}$, respectively. After deposition, the sample was annealed for $2 \mathrm{~h}$ at 800 $^{\circ} \mathrm{C}$ in 300 mbar of ambient oxygen environment and cooled at 10 $^{\circ} \mathrm{C} / \mathrm{min}$ rate. The TmIG layer was grown by maintaining the substrate temperature at 750 $^{\circ} \mathrm{C}$, laser fluence at $1 \mathrm{~J} / \mathrm{cm}^2$, with a repetition rate of $6 \mathrm{~Hz}$ in 0.26 mbar of oxygen partial pressure. Post deposition, the prepared TmIG film was in-situ annealed at same growth temperature for 25  mins at 100 mbar oxygen pressure followed by cooling at 5 $^{\circ} \mathrm{C} / \mathrm{min}$. During deposition the substrate to target distance was kept $5 \mathrm{~cm}$ for both the films. The bilayer samples were prepared by following the same growth condition as the individual single layers. For the bilayer samples, each FiM layer was first deposited and then annealed in similar conditions as of its corresponding single reference layer samples and then the subsequent layer was deposited followed by its post-annealing. The details of prepared sample structure are mentioned in Table I. The thickness of the corresponding layer is mentioned in the brackets. The Pt layer has been prepared via \textit{dc} magnetron sputtering in a high vacuum multi-deposition chamber manufactured by Mantis Deposition Ltd., UK.

The growth quality and thickness of the prepared films were investigated by $\mathrm{X}$-ray diffraction (XRD). High-resolution transmission electron microscopy (HR-TEM) has been performed in this study to verify the epitaxy of deposited films. The saturation magnetization value has been taken from the superconducting quantum interference device (SQUID) magnetometry data. The ferromagnetic resonance (FMR) and spin pumping induced ISHE measurements were performed using a coplanar waveguide (CPW) based setup. The sample was placed upside down on the CPW. FMR measurements were carried out in the 3-12 GHz of frequency range with 25 mW microwave power. The ISHE measurements were carried out at 7 GHz \textit{rf} frequency by connecting a nanovoltmeter at the opposite edges of the sample. The measurement details are mentioned in our previous work \cite{singh2020large}.

\begin{table}
\caption{Details of the sample structure studied in this work }

\begin{tabular}{|c|c|}
\hline Sample  & Sample \\
name & structure\\
\hline S1 & GGG(111)/YIG(100 nm)/Pt( $(5 \mathrm{~nm})$ \\
\hline S2 & GGG(111)/TmIG(30 nm)/Pt(5 nm) \\
\hline S3 & GGG(111)/YIG(100 nm)/TmIG(30 nm)/Pt(5 nm) \\
\hline S4 & GGG(111)/TmIG(30nm)/YIG(100 nm)/Pt(5 nm) \\
\hline
\end{tabular}
\end{table}

\begin{figure}[ht]
	\centering
	\includegraphics[width=0.5\textwidth]{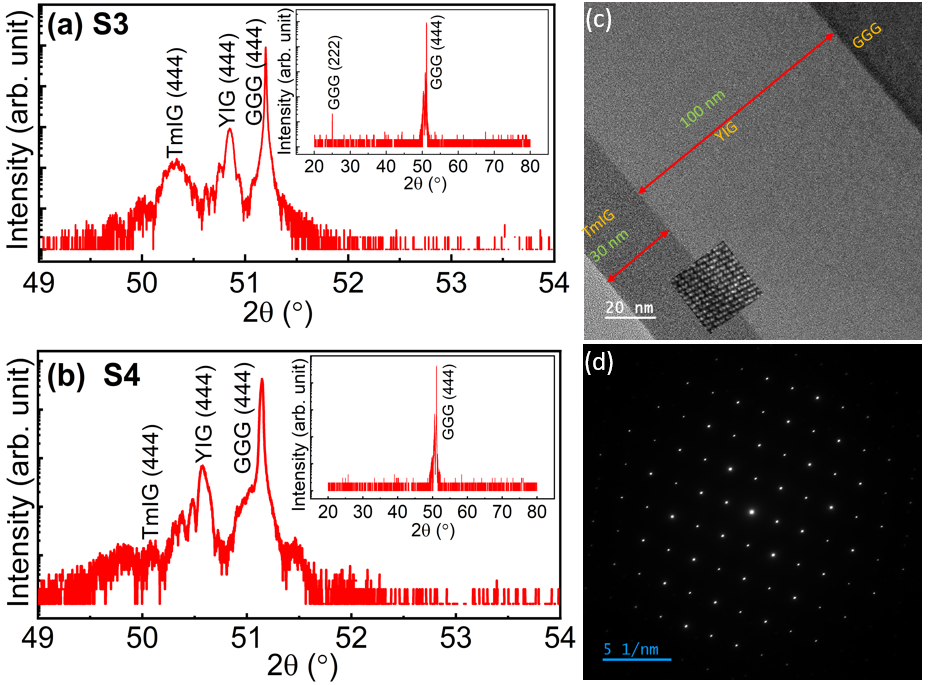}
	\caption{XRD patterns of (a) S3 and (b) S4 samples (the corresponding insets show the XRD pattern for the 2$\theta$ range $20^{\circ}$-$80^{\circ}$). HRTEM image and SAED pattern for sample S3 are shown in (c) and (d), respectively. The inset shows the magnified part of YIG and TmIG interface. }
	\label{fig:Fig._1}
\end{figure}

The prepared samples were structurally characterized by XRD to confirm the phase and growth quality. The XRD pattern of S3 and S4 samples are shown in Fig. 1(a-b). The XRD pattern of S1 and S2 is given in the supplementary file. The YIG and TmIG lattice parameters are very close to that of the substrate GGG which ensures the epitaxial growth of the structure. The observed diffraction peaks were indexed with the corresponding crystal plane (h k l) values, and it is evident from the analysis that the growth of prepared samples is along the (111) direction. The absence of any additional peaks other than the peaks corresponding to YIG, TmIG, and GGG in the patterns ensures the phase purity of the grown structures (shown in the corresponding insets). 
In Fig. 1(a), the peak corresponding to YIG layer is dominating over the TmIG peak as the YIG thickness is high. By analyzing the XRD peak at (444) reflection yields a cubic lattice parameter of YIG in S1 is 12.46 \AA, which is comparable to 12.38 \AA  for the bulk YIG \cite{gilleo1958magnetic}. Moreover, the lattice parameters of YIG and TmIG are estimated for all the samples. The YIG lattice parameter in S3 and S4 is 12.43 \AA  and 12.48 \AA, respectively. Likewise, the lattice parameters of TmIG in S2, S3, and S4 are 12.45 \AA, 12.54 \AA, and 12.60 \AA, respectively. It is to be noted that, the lattice parameter has changed in S3 and S4 samples for both YIG and TmIG with respective to their single layer i.e., S1 and S2.This indicates the presence of strain in the films which may have the impact on the physical properties of the samples.

  The interface is further explored by cross-sectional HRTEM for the sample S3. The zoomed-in image [shown in the inset of Fig. 1(c)] reveals the epitaxial growth of YIG and TmIG. These high-quality images provide clear evidence of the well-defined interface and crystalline structure of the film showing in-plane lattice matching with the substrate. The thicknesses of YIG and TmIG were found to be 100 nm and 30 nm respectively. 
Notably, no defects or misalignment in lattice planes were observed in the HRTEM image shown in Fig. 1(c). Fig. 1 (d) shows the selected area electron diffraction (SAED) pattern of sample S3, which confirms the single crystalline nature. The validation provided by these HRTEM images and SAED pattern is crucial evidence, confirming the successful fabrication of the thin films with sharp interface. 

The spin dynamics properties were investigated by FMR measurements at room temperature. Fig. 2 (a-d) show the FMR spectra of the prepared samples at different frequencies ($f$) ranging from 3- 12 GHz. Distinct FMR peaks were observed for YIG and TmIG in all samples. This confirms the room temperature magnetic phase of the prepared samples. A shift in $H_{res}$ corresponding to YIG and TmIG in the S3 and S4 samples compared to S1 and S2 samples is observed. This could be attributed to the interfacial exchange coupling between YIG and TmIG, similar to other multilayered systems \cite{li2020coherent}.   
\begin{figure}[ht]
	\centering
	\includegraphics[width=0.5\textwidth]{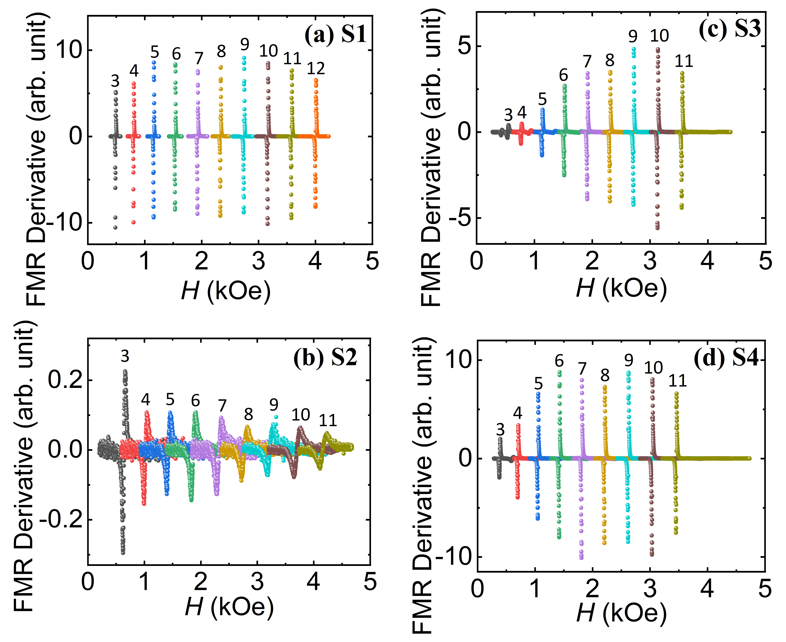}
	\caption{FMR spectra of (a) S1 (b) S2 (c) S3 and (d) S4 samples at different frequencies. }
	\label{fig:Fig._2}
\end{figure}

\begin{figure*}[ht]
	\centering
	\includegraphics[width=0.8\textwidth]{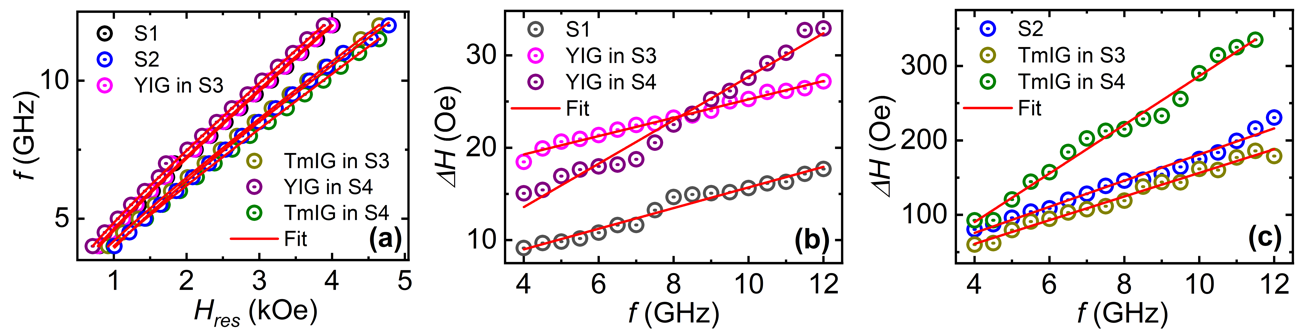}
	\caption{(a) $f$ vs $H_{res}$ plot for all the samples. $\Delta H$ vs $f$ plot for (b) YIG and (c) TmIG for respective resonance in FMR signals with corresponding fittings.}
	\label{fig:Fig._3}
\end{figure*}

Further, the FMR signals were fitted by Lorentzian function to obtain line width $(\Delta H)$ and resonance field ($H_{res}$) at each frequency. Later, the damping analysis is carried out for all the samples by plotting $f$ vs $H_{res}$ and $\Delta H$ vs $f$ (shown in Fig. 3 ). In case of S1, the damping value is estimated considering the uniform (n=0) mode. Here, for the bilayer samples, the damping analysis is done for the respective resonance fields of YIG and TmIG.

The plotted data in Fig 3 (a) were fitted by the Kittel equation \cite{kittel1948theory},

\begin{equation}
f=\frac{\gamma}{2\pi}\sqrt{(H_{res}+H_{K})(H_{res}+4\pi M_{eff}+H_{K})}
\label{Kittel equation}
\end{equation}

where, $\gamma\left(=\frac{g \mu_B}{\hbar}\right.$ ) is gyromagnetic ratio ( $g$ is Lande $g$-factor, $\mu_B$ is Bohr magneton), $H_K$ is in-plane anisotropic field), $4 \pi M_{e f f}\left(=4 \pi M_s+\frac{2 K_S}{M_s t_{F M}}\right)$ is the effective demagnetizing field $\left(K_S, M_s\right.$, and $t_{F M}$ are perpendicular surface anisotropy constant, saturation magnetization and thickness of the magnetic layer, respectively). Afterwards, the Gilbert damping constant ($\alpha$) values were estimated by fitting the plot in Fig 3(b) and (c) by the equation

\begin{equation}
\Delta H=\Delta H_{0}+\dfrac{4\pi\alpha f}{\gamma} .
\label{non linear equation fit}
\end{equation} 

The obtained values of $g$, $M_{\mathrm{s}}$ and $\alpha$ (tabulated in Table II) are in good agreement with the existing literature \cite{haertinger2015spin,li2020impact}. An increase in $\alpha$ value is observed for both YIG and TmIG in S3 and S4 samples compared to the $\mathrm{S} 1$ and $\mathrm{S} 2$ samples.

The ISHE measurements were carried out by a FMR based setup. The observed voltage signal corresponding to the FMR resonance is shown in Fig 4 (a-d) for the prepared samples. In general case, for ISHE two layers are required, one is the source for spin current i.e., the magnetic layer and the other one is the spin sink i.e., the high spin orbit coupling (HS) material. In this study, the HS layer is the Pt layer where the spin current source is the YIG/TmIG bilayer in the samples S3 and S4. Interestingly, here we have observed ISHE voltage at two distinct FMR resonance field positions corresponding to the resonances and concomitant spin currents mainly excited from different layers, i.e., YIG and TmIG layers in both S3 and S4 samples as shown in Fig. 4.

In order to extract the spin pumping voltage from the rectifications such as anomalous Hall effect (AHE) and anisotropic magneto resistance (AMR), the voltage was measured at different value of the angle ($\phi$) at a step of $5^{\circ}$ in the range of $0^{\circ}$ to $360^{\circ}$ at a constant frequency of 7 GHz. We deliberately used higher frequency to avoid the 3 magnon splitting phenomenon which in general takes place at lower frequencies \cite{kurebayashi2011controlled}. Here $\phi$ is defined as the angle between the direction of $H$ and the perpendicular direction to contacts for voltage measurement. The measured voltage is fitted by the Lorentzian equation. The symmetric ($V_{sym}$) and antisymmetric ($V_{asym}$) contributions of the voltage were extracted. The obtained $V_{sym}$ and $V_{asym}$ values were plotted as a function of angle $\phi$ (shown in Fig. 5). It is to be noted that the antisymmetric component is almost negligible compared to the symmetric component. This clearly indicates the dominance of the spin pumping induced ISHE in the samples.

Moreover, to quantify the spin pumping voltage and other rectification effects the estimated $V_{sym}$ and $V_{asym}$ values in Fig. 5 were fitted by the given equations (3) and (4), respectively \cite{iguchi2017measurement}.
\begin{table*}
\caption{Fitted parameters}
\begin{tabular}{|c|c|c|c|c|c|c|}
\hline \multirow[t]{2}{*}{ Sample } & \multirow[t]{2}{*}{ S1(GGG/YIG/Pt) } & \multirow[t]{2}{*}{ S2(GGG/TmIG/Pt) } & \multicolumn{2}{|c|}{ S3(GGG/YIG/TmIG/Pt) } & \multicolumn{2}{|c|}{ S4(GGG/TmIG/YIG/Pt) } \\
\hline & & & YIG & $\mathrm{TmIG}$ & YIG & TmIG \\
\hline $\mathrm{\alpha} \times 10^{-3}$ & $0.51 \pm 0.07$ & $17.00 \pm 0.03$ & $1 .10 \pm 0.03$ & $17.00 \pm 0.04$ & $2.60 \pm 0.01$ & $30.00 \pm 0.38 $ \\
\hline$V_{sp}(\mu V)$ & $180.0 \pm 7.8$ & $0.06 \pm 0.01$ & $18.0 \pm 0.7$ & $1.00 \pm 0.01$ & $220.00 \pm 0.52$ & $0.16 \pm 0.12$ \\
\hline$V_{\text {AHE }}^{\text {asym }}(\mu V)$ & $17.0 \pm 1.5$ & $0.010 \pm 0.004$ & $0.94 \pm 0.07$ & $0.16 \pm 0.07$ & $3.20 \pm 0.82$ & $(0.40 \pm 0.02) \times 10^{-2}$ \\
\hline$V_{AMR}^{\perp} (\mu \mathrm{V})$ & $12.0 \pm 1.0$ & $0.04 \pm 0.01$ & $11.0\pm0.8$ & $0.58 \pm 0.01$ & $1.10 \pm 0.16$ & $0.17 \pm 0.15$ \\
\hline$V_{AMR}^{\|}(\mu \mathrm{V})$ & $0.62 \pm 0.04$ & $(0.40 \pm 0.04) \times 10^{-2}$ & $1.40\pm0.38$ & $0.110 \pm 0.002$ & $1.40 \pm 0.14$ & $0.050 \pm 0.006$ \\
\hline$g^{\uparrow\downarrow}_{eff}$ & $4.78 \times 10^{18}$ & $2.39 \times 10^{17}$ & $1.56 \times 10^{17}$ & $1.12 \times 10^{17}$ & $1.24 \times 10^{18}$ & $1.76 \times 10^{18}$ \\
\hline
\end{tabular}

\end{table*}

\begin{equation}
\begin{aligned}
{V_{sym}= V_{sp}cos^3(\phi)+V_{AHE}\ cos(\theta)}cos(\phi)\\
    + V_{sym}^{AMR \perp} cos (2\phi)cos(\phi)\\
    + V_{sym}^{AMR ||}sin(2\phi)cos(\phi)
    \end{aligned}
\end{equation}
\begin{equation}
\begin{aligned}
 V_{asym}= V_{AHE}\ sin (\theta) cos(\phi) + 
    \\V_{asym}^{AMR \perp} cos (2\phi)cos(\phi)+ 
   \\ V_{asym}^{AMR ||}sin(2\phi)cos(\phi)
   \end{aligned}
 \end{equation}

where $V_{sp}$ and $V_{AHE}$ are voltages due to spin pumping and the anomalous Hall effect. Furthermore, $V_{asym, sym}^{AMR\parallel}$ and $V_{asym, sym}^{AMR\perp}$ are the parallel and perpendicular components of the AMR voltage, respectively. $\theta$ is the angle between the electric and magnetic fields of the microwave which is $90^{\circ}$.

\begin{figure}[ht]
	\centering
	\includegraphics[width=0.5\textwidth]{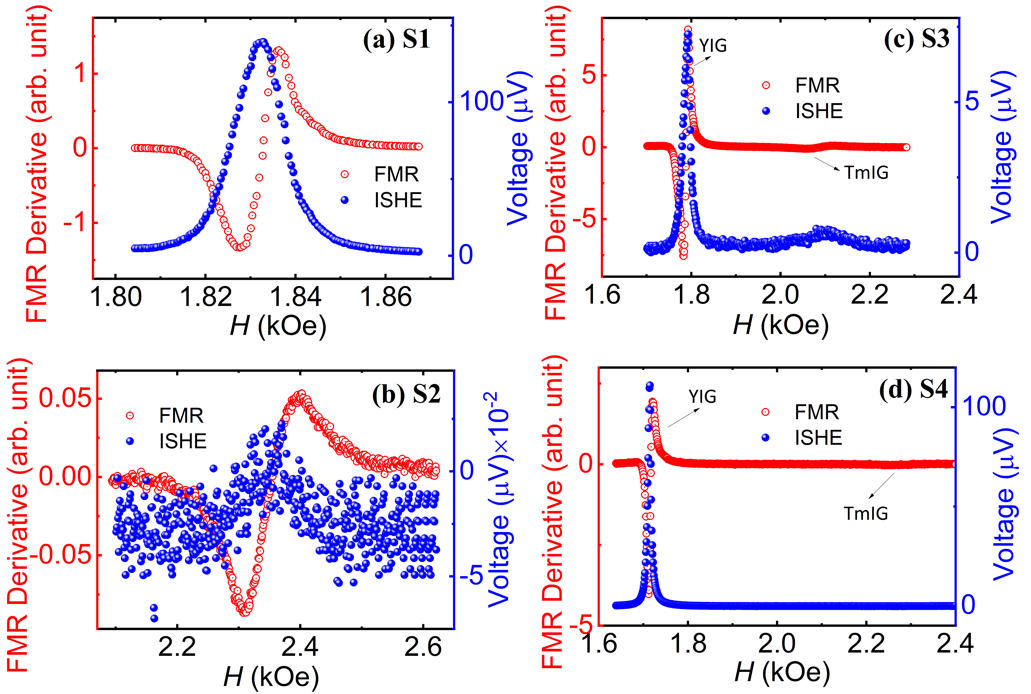}
	\caption{(a)FMR and ISHE spectra for (a) S1, (b) S2, (c) S3 and S4 samples at 7 GHz.}
	\label{fig:Fig._4}
\end{figure}

The extracted value from fitting is tabulated in Table II. The $V_{A M R}^{\|, \perp}$ component is calculated by the following formula\cite{conca2017lack}

\begin{equation}
\begin{aligned}
V_{A M R}^{\|, \perp}=\sqrt{\left(V_{s y m}^{A M R \|, \perp}\right)^2+\left(V_{a s y m}^{A M R \|, \perp}\right)^2}
    \end{aligned}
\end{equation}

\begin{figure}[ht]
	\centering
	\includegraphics[width=0.5\textwidth]{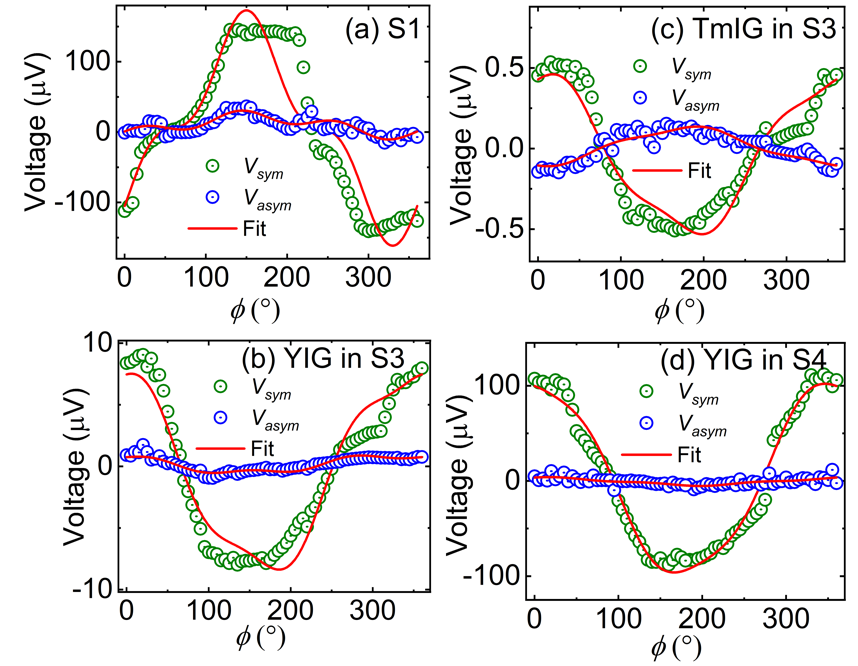}
	\caption{Angular dependence of $V_{sym}$ and $V_{asym}$ with corresponding fits for (a) S1, (b) YIG in S3, (c) TmIG in S3, and (d) YIG in S4.}
	\label{Fig._5}
\end{figure}

From Table II, it is further confirmed quantitatively that the spin pumping voltage $V_{sp}$ is the dominating contribution to the measured ISHE voltage. The power-dependent data (shown in supplementary file Fig. 2) further ensure the spin pumping dominance in all the samples. It is observed that the $V_{sp}$  has decreased by one order for YIG in S3 sample as compared to the S1 sample. This is expected as in S3 sample, the spin current generated at YIG layer has to pass through the TmIG layer before converting to charge current at Pt layer. In this process, there may be dissipation of the spin angular momentum which led to the decrease in the spin pumping voltage. The $V_{sp}$  is almost 20 times larger for TmIG in S3 as compared to the S2 sample. The observed enhanced $V_{sp}$  for TmIG in S3 and S4 as compared to the S2 sample can be ascribed to the interfacial exchange coupling. The presence of interlayer exchange coupling in magnetic bilayer systems has already been shown to change the amplitudes of different ferromagnetic resonance modes due to the dynamic exchange field at the interface \cite{qin2018exchange,klingler2018spin}. As observed from the FMR data, the coupling of the magnon current at the interface of YIG and TmIG led to the enhancement of the $V_{sp}$  for TmIG. Interestingly, the $V_{sp}$  is maximum for YIG in S4 sample. This may be ascribed to the interfacial exchange coupling between YIG and TmIG.

In this context, in order to quantify the spin current propagation, effective spin mixing conductance (g$^{\uparrow\downarrow}_{eff}$) is evaluated by the following expression using the obtained damping constant value \cite{thiruvengadam2022anisotropy}.

\begin{equation}\label{eq8}
%\begin{aligned}
g^{\uparrow\downarrow}_{eff}=\frac{4 \pi \Delta \alpha M_s t_{F M}}{g \mu_B}
    %\end{aligned}
\end{equation}

where $M_{S}$, $t_{FM}$, and $\Delta$$\alpha$ are the saturation magnetization, thickness of magnetic layer and change in Gilbert damping from bilayer to single layer films, respectively. These g$^{\uparrow\downarrow}_{eff}$ values (of the order of 10$^{18}$ m$^{-2}$) are well matched with the existing literature\cite{haertinger2015spin}. Hence, the interfacial exchange coupling led to the enhancement of spin pumping for YIG and TmIG layer \cite{liu2023strong}.

We have studied spin pumping and ISHE for garnet/Pt systems. The damping analysis exhibits an enhancement of $\alpha$ in the bilayer garnet sample.  The measured ISHE voltage for YIG/Pt layer is around 180 $\mu$V where the major contribution is from spin pumping. A decrease in the ISHE voltage is observed by one order i.e., 17$\mu$V in S3 which is attributed to the presence of TmIG layer playing as a hindrance to the transfer of angular momentum from the YIG layer. Whereas, an increase in spin pumping voltage for YIG in S4 and TmIG in S3 is observed as compared to their respective single layers. This may be attributed to the interfacial exchange coupling between YIG and TmIG. Moreover, further study can be carried out to have an insight to interface exchange coupling and the effects of spin pumping on magnon-magnon interactions.

\section*{Acknowledgement}
S.B., A.S., K. S. R., P.G., and A.M. thank the Department of Atomic Energy, Department of Science and Technology, Science and Engineering Research Board (Grant No. CRG/2021/001245), Government of India, Chanakya Post-doctoral fellowship, i-Hub quantum technology foundation (Sanction Order No. I-HUB/PDF/2022-23/04) for providing financial support.  Work from J. L. and A. H. with respect to the data analysis and discussion, as well as manuscript preparation has been supported by the U.S. Department of Energy, Office of Science, Basic Energy Sciences, Materials Sciences and Engineering Division through Contract No. DE-SC0022060.

\section*{DATA AVAILABILITY}
The data that support the findings of this study are available from the corresponding author upon request.

\section*{REFERENCES}
\bibliographystyle{apsrev4-2}
\bibliography{manuscript}

\end{document}